\documentstyle[emulateapj,psfig]{article}

\begin{document} 
\newcommand{\be}{\begin{equation}}
\newcommand{\ba}{\begin{eqnarray}}
\newcommand{\ee}{\end{equation}}
\newcommand{\ea}{\end{eqnarray}}

\title {Detecting the Gravitational Redshift of Cluster Gas }

\author
{Tom Broadhurst \& Evan Scannapieco}
\affil
{1) European Southern Observatory, Munich, Germany. \\
2) Department of Astronomy, University of California, Berkeley, CA 94720}

\begin{abstract}
  
We examine the gravitational redshift of radiation emitted from within
the potential of a cluster. Spectral lines from the intracluster medium 
(ICM) are redshifted
in proportion to the emission-weighted mean potential along the line
of sight, amounting to $\approx 50$ km/s at a radius of 100 kpc/h, for
a cluster dispersion of 1200 km/s. We show that the relative redshifts
of different ionization states of metals in the ICM provide a unique probe of
the three-dimensional matter distribution.  An examination of the reported peculiar
velocities of cD galaxies in well studied Abell clusters reveals they
are typically redshifted by an average of $\sim +200$ km/s. 
This can be achieved by gravity with the addition of a steep
central potential associated with the cD galaxy. Note that in general 
gravitational redshifts cause a small overestimate of the
recessional velocities of clusters by an average of $\sim$ 20 km/s.

\end{abstract}
	


\section{Introduction}

  Recent N-body work has revealed that the central mass profile of
cluster-sized dark matter halos in pure Cold Dark Matter (CDM)
simulations are relatively flat inside a radius of of $\sim 400$ kpc,
asymptoting to $r^{-1}$ (Navarro, Frenk, \& White 1997, hereafter NFW;
Moore et al.\ 1998).  This prediction, although distinct, is not
easily examined with lensing observations. The distortion of images by
lensing is a powerful way of locating dark matter but is insensitive
to the gradient of the distribution, particularly in the central
region which is subject to a degeneracy (Schneider \& Seitz 1995;
Kaiser 1995). Thus the flattening of the slope to $r^{-1}$ cannot be
distinguished in practice from a singular isothermal profile from
lensing distortions alone (e.g.\ Clowe et al.\ 1998). In principle,
the magnifications of lensed sources are more sensitive to the
projected mass profile,
but their use requires knowledge of both
the source redshifts and intrinsic unlensed properties to be reliable
(Broadhurst, Taylor, \& Peacock 1995; Taylor et al.\ 1998). Of course
all lensing signals relate only to the projected mass distribution and
thus provide only indirect information along the line of sight.

Here we evaluate a means of exploring the distribution of matter in
clusters that complements and extends the information available from
lensing.  The gravitational potential of a cluster not only deflects
light passing nearby, but imprints a gravitational redshift on light
emitted by the cluster itself. The many X-ray emission
lines in the metal-rich intracluster medium provide a suitable tracer
of this redshift.  Since the line emission is generated by a two body
process, the velocity information is weighted by the square of the gas
density along a given sight-line, and is quite sensitive to the mass
profile in the central region of the cluster.  Furthermore,
temperature gradients in the gas cause different lines to be emitted
at different distances along a line of sight through the cluster, and
may thus provide a three-dimensional probe of the matter profile.

While X-ray spectroscopy has been difficult in the past, it is
advancing to the point now where this effect may be marginally
detected. In the near future, an order of magnitude increase in
resolution and effective collecting area will be achieved with
improved techniques.

The structure of this paper is as follows. In \S2 we calculate the
form of the gravitational redshift for simple mass profiles.
In \S3 we discuss the properties of more realistic cluster profiles and 
the possibility of using multiple metal lines to study the distribution of
matter along the line of sight.  Finally in \S4 we make some comments
regarding existing optical data and prospects for the near future.

\section {Gravitational Redshift}

\subsection{Simple Analytical Model}

The simplest useful model of a cluster
is an isothermal sphere in which the density 
$n(r)$ is proportional to  $r^{-2}$ for radii less than a truncation 
radius $R_0$, and $n(r)=0$ for $r > R_0$, so that the gravitational 
potential is given by
\be
\phi(r) = 
\cases {
 2 \sigma^2 [\ln(r/R_0) -1]  &  $r<R_0$ \cr
 -2 \sigma^2 R_0/r   &  $r>R_0$, \cr}
\label{eq:phiiso}
\ee
where $\sigma$ is the velocity dispersion. 
The gravitational redshift of light from point sources such as galaxies
 located within the potential is given simply by
\be
\Delta \nu(r) = \nu_0 \frac{\phi(r)}{c^2},
\label{eq:nushiftpoint}
\ee
where $\nu_0$ is the rest-frame emission frequency.
Spectral line emission from the gas, however, is  
proportional to $n_e(r) n_{\rm ion}(r)$ and
hence assuming
the metals are distributed like the electrons we may simply 
weight the potential by $n^2(r)$ along 
lines of sight to compute the average redshift.
In this case, the frequency of the emission line will be shifted 
by an amount
\be
\Delta \nu(x) =
\frac{\nu_0}{c^2}  
\frac{\int_{0}^{R(x)} dl \,\phi((x^2+l^2)^{1/2})) n((x^2+l^2)^{1/2})^2}
     {\int_{0}^{R(x)} dl \, n((x^2+l^2)^{1/2})^2},
\label{eq:nushift}
\ee
where $l$ is the distance along the line of sight
and $R(x) \equiv \sqrt{R_0^2-x^2}$.  

For a singular isothermal sphere then,
\ba
z(x) &=& -\frac{\Delta \nu(x)}{\nu_0} = \frac{2\sigma^2}{c^2} 
\left[
\frac{1}{2} 
-\frac{L(\theta_x) - \theta_x + \theta_x \ln \cos \theta_x}
    {\theta_x + \sin \theta_x \cos \theta_x} 
\right] \nonumber \\
&\approx& \frac{2\sigma^2}{c^2} 
\left[\frac{3}{2} - \ln\left( \frac{2x}{R_0} \right) \right],
\label{eq:dzsec1}
\ea
where $\theta_x \equiv \arccos(x/R_0),$
$L(\theta_x) \equiv - \int_0^{\theta_x} d\theta \ln (\cos \theta)
$ is Lobachevskiy's function, and
the approximation is valid for small impact parameters.
The velocity dispersion is only a weak function of cosmology 
(Eke, Cole, \& Frenk 1996) and can be estimated as 
$\sigma=730 \, M_{14}^{1/3} (1+z_c)^{1/2} \,{\rm km/s},$
where $M_{14}$ is the mass of the collapsed halo in units of $10^{14} h^{-1}$
solar masses and $z_c$ is the redshift of collapse. 
Finally, the relation between the total mass and the velocity dispersion 
($\frac{G M}{R_0}=2 \sigma^2$) gives us the truncation radius
$R_0=400 \, M_{14}^{1/3} (1+z_c)^{-1} \,{\rm kpc/h}.$

In the top panel of Figure \ref{fig:rofz}, we plot the redshift as a
function of radius for three clusters with $M_{14}=$ 1.5, 3, and 6,
and $z_c=0.5$.  Here we see that while the gravitational redshift of
emission lines near the edge of the cluster is quite small, the
$r^{-4}$ weighting boosts the effect significantly along lines of
sight near the center of the cluster, reaching $\sim 10 \%$
of the velocity dispersion within the inner $10$ to $50$ kpc.  

\subsection{Simple Numerical Model}

To better treat the interesting core region of the
cluster, let us consider a simple nonsingular model (Binney \& Tremaine
1987) derived by numerically integrating the equation of hydrostatic
support of an isothermal gas 
\be 
\frac{d}{d \tilde r} \left(\tilde r^2
\frac{d \ln \tilde \rho}{d \tilde r} \right) = - 9 \tilde r^2 \tilde \rho,
\ee 
where the density and radius have been rescaled such that $\tilde
\rho \equiv \rho / \rho_0$ and  $\tilde r \equiv r/ r_0$, where $\rho_0$
is the central density of the cluster and $r_0$ is the King radius,
$r_0 \equiv \sqrt{ \frac{9 \sigma^2}{4 \pi G \rho_0}}$.  Integrating
outwards from the boundary conditions $\tilde \rho(0)=1$ and
$\frac{d \tilde \rho}{d \tilde r}=0$ determines the density and
gravitational potential at all radii, 
and we truncate the cluster at $10 r_0$.  

Again we can relate mass and velocity dispersion 
and solve for $r_0$ as a function of mass and redshift, which gives
$r_0 = 38 \, M_{14}^{1/3} (1+z_c)^{-1} \,{\rm kpc/h}$.
We then use $\phi(r)$ and $n(r)$ to numerically
integrate Eq.\ \ref{eq:nushift} to obtain the mean redshift of
lines as a function of impact parameter $x$.
In the center panel of Figure \ref{fig:rofz} we plot our 
results for three clusters of mass $M_{14}=$ 1.5, 3, 6 and $z_c=0.5$,
as the results for a nonsingular sphere in which the potential
associated with a central cD galaxy has been simply added to the numerical
solution, modeled as a point mass with $2\%$ of the cluster mass.
For reference, we also plot $\phi(r)$ for the singular, nonsingular,
and nonsingular plus point mass models in the bottom panel.

In this figure we see that the singular and nonsingular isothermal
models have much the same $z(x)$ profiles outside the core radius, but
the presence of a core creates a sharp turnover at small radii,
allowing these two potentials to be easily distinguished. Note also
that the sharp increase in $\phi(r)$ at very small radii in the core
plus point mass model leads to only a $\sim$ 10 km/s increase in the
observed line redshifts, due to averaging along the line of sight.

\section{NFW Model and Temperature Gradients}

\begin{figure}
\centerline{\psfig{file=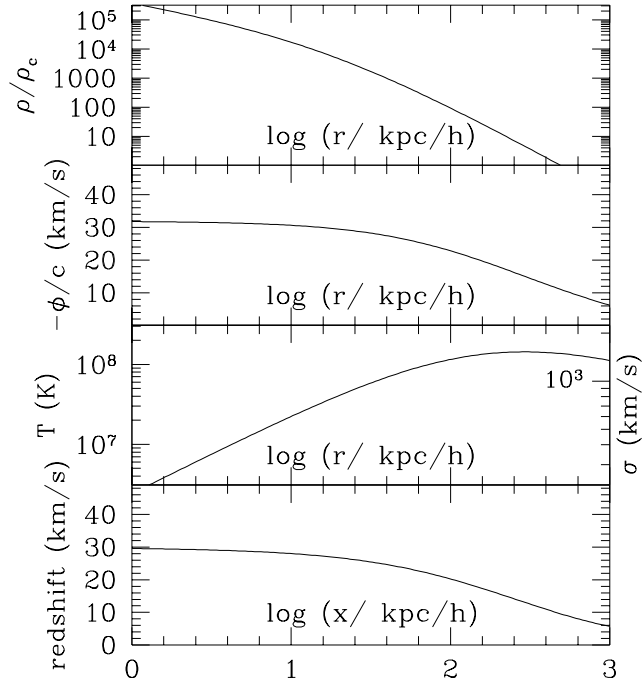,width=4.1in}}
\caption{NFW model for a $M_{14}=3$, $z_c=0.5$ cluster.  The
upper panels show the density profile, cluster potential, and
temperature (velocity dispersion) as a function of radius.  The lower
panel shows the ``naive'' gravitational redshift as a function of
impact parameter as predicted by Eq.\ \protect\ref{eq:nushift}.}
\label{fig:nfw}
\end{figure}

The models adopted in \S2 were intended only as illustrative,
and by no means represent the full physics of the cluster dark
matter and gas. 
The evolution of the dark matter in clusters is complex, many showing
clear evidence for recent mergers (Zabludoff \& Zaritsky 1995; Bird 1995; 
Bird, Davis, \& Beers 1995; Bliton et al.\ 1998; Burns et al. 2000). 
For many quiescent clusters, however, 
the radial profile of simulated
clusters in CDM models are given by a universal density profile 
as shown in NFW:
\be
\rho \propto \frac{1}{r/r_s(1+r/r_s)^2}, 
\ee where $r_s = \frac{760}{C}
{\rm kpc}/h M_{14}^{1/3} (1 + z_c)^{-1}$, $C \approx 3.6 (1 +
z_c)$, and the profile is truncated at $r/r_s=C$. 
Figure \ref{fig:nfw} illustrates one such distribution
with $M_{14}=3$ and $z_c=0.5$.
Here we see that the steepness of the density profile
and the flattening of the  potential conspire to give a 
gravitational redshift that closely traces $\phi(r)$
and is very different from the singular isothermal model.  
Unlike the simple models considered in \S2, however, the 
redshift estimate provided by Eq.\ \ref{eq:nushift} does not
tell the full story, as it does not account for the presence
of a temperature gradient.

In general the gravitational redshift will not only be a function of
gas density along the line of sight, but of the temperature profile as well.
The ionization potential of each emission line will be
strongly peaked about its ionization temperature (e.g.\ Zombeck 1990),
and hence an emitted line will be  confined to within a hollow shell
in the cluster.  Thus, with
measurements of different X-ray emission lines, one can hope to
simultaneously solve for the potential and the temperature profile,
gaining three-dimensional information. This must be taken into account in a NFW
potential and will be especially important for clusters containing
cooling flows.  In this case the temperature gradient may exceed
two orders of magnitude, so that for example the excitation of iron may
range from FeX to FeXXVI (Fabian 1994; Churazov et al.\ 1998). 

\begin{figure}
\centerline{\psfig{file=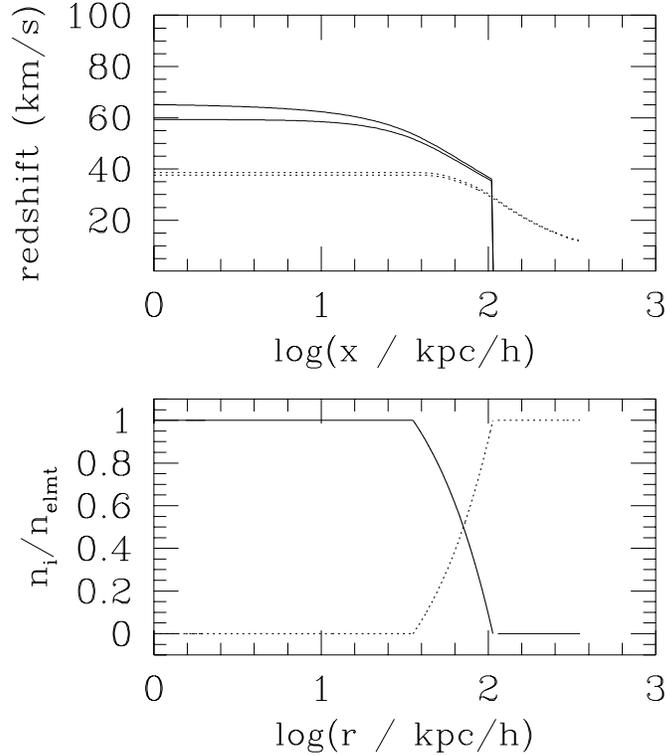,width=4.1in}}
\caption{ Gravitational redshift for the $M_{14} = 3$ model in Figure 2
in which the effects of temperature variations are modeled simply by
taking the fractional ionization between two ionization states to vary
linearly from $1r_0$ to $3r_0$.  The lower ionization state is given by 
the solid lines and the higher by the dotted lines.  {\em Top:} Observed
gravitational redshift of the lines from the two ions
as a function of impact parameter of the line of sight.
 {\em Bottom:}
The fractional number densities of each ionization
state as a function of radial distance.    }
\label{fig:temp}
\end{figure}

A simple illustration of the additional information afforded by
temperature variations is given in Figure \ref{fig:temp}.  Here the
fractional ionization between two ionization states $X_{i+1}$ and
$X_i$ is allowed to vary linearly from $1r_0$ to $3r_0$, in the
potential given in \S2.2, with $M_{14} = 3$ and both with and without
a point mass.  We interpret this model as a cooling-flow configuration
in which the higher ionization state cannot exist within the core
region of the cluster where the temperature drops much lower than its
ionization potential.  Here we see that the $x$ beyond which $X_i$ is
completely ionized to the higher state is marked by an abrupt cutoff
in line emission.  The $x$ at which $X_{i+1}$ completely recombines to
the lower state is marked by a leveling off in the redshift as a
function of $x$, accompanied by a continuing increase in redshift as a
function of $x$ in the lines from the lower ionization state.  Along
lines of sight such that $r_0 \leq x \leq 3 r_0$, which at $\ell = 0$ pass
through the partially ionized region, we see that the lines are offset
from each other, providing information as to the width of the
transition region along the line of sight.  This can be compared to
the width of the partially ionized region in the $x$ direction to
determine the sphericity of the cluster.

\section{Observations}

\subsection{Redshift of Central Star Light}

While X-ray observations are the cleanest approach, the redshift of
spectral features from starlight emitted from within the potential may
also be studied as first suggested by Cappi (1995).
Carefully studied massive clusters with central cD galaxies can be
found in Sharples, Ellis, \& Gray (1988) and
the sample of Zabludoff et al.\ (1993) and are listed below
in Table \ref{tab:anom}, excluding clusters with dispersions below 700
km/s for which a significant redshift is not expected (Eq.\
\ref{eq:nushiftpoint}).  Interestingly, all these peculiar velocities
but one are redshifted.  In this table, the error listed in $\Delta
V_{\rm cD}$ is the error in the measurement of the cD galaxy alone
($\sigma_{\rm cD}$) while the error listed in the following column is
the error of the systemic velocity of the cluster ($\sigma_{\rm
sys}$).  The total error on the peculiar cD velocity is thus
$\sigma_{\rm gr} = (\sigma_{\rm cD}^2+ \sigma_{\rm sys}^2)^{1/2}$.  In
some cases the clusters have well known cooling flows which we have
noted below as the central potential may have been deepened by the 
action of a substantial flow over time.
Here we use the values from
the ROSAT study of Peres et al.\ (1998), quoting results obtained
using the Position Sensitive Proportional Counter (PSPC) when multiple
values were given.

Substructure has been sought thoroughly for the above clusters and can
always be identified at some level (Hill \& Oegerle 1993, 1998;
Oegerle \& Hill 1994; Bird 1994; Kriessler \& Beers 1997) but is not 
enough to account for the systematic redshift. 
Note we have excluded the cluster A85 for which significant clustering
may be responsible for a large peculiar redshift of +1100 km/s (Bird
1994).

Weighting the sum by the variance, we arrive at an average $\Delta V_{cD}$
of $260 \pm 58$ km/s. 
While this analysis is by no means exhaustive and does not attempt any
correction for substructure, we may feel confident
that the sign of this effect is real, namely a redshift, and the rough
order of magnitude is not unreasonable.  
From Eq.\ \ref{eq:nushiftpoint} we can relate this redshift to the
cluster potential.
The error weighted ratio of $\Delta V_{\rm cD}$ and the velocity dispersion 
is $(60 \pm 20) \sigma^2/c$, or $340 \pm 110$ km/s for $\sigma = 1300$ km/s,
which, while somewhat higher than our predicted
redshifts, is nevertheless reasonable and may suggest that the
form of the central potential is steeper than the slow 
logarithmic radial variation of a purely isothermal potential.

\subsection{Future Prospects and Observational Issues}

While the systematic redshift of cD galaxies examined above provides
a tantalizing suggestion of cluster gravitational redshifts, the 
possibility of subtle substructure makes studying optical emission from 
cluster galaxies difficult to apply accurately.  At best such studies are 
limited by the finite number of bright cluster members for measuring accurate 
velocities.

Thus the X-ray spectroscopic approach outlined in \S2 and \S3 will provide the
most straightforward method for studying gravitational redshifts.
Note that this requires both high spectral resolution to determine
the shift of the line centroids, and fair spatial resolution to
obtain radial information. 
Little information can be learned from a purely spectral analysis, however,
as the overall cluster-integrated line profile is indistinguishable from a 
thermal profile with a slightly broadened width.

The exploration of this effect may be marginally possible with the
current generation of X-ray satellites. The High-Energy Transmission
Grating on the Chandra Satellite, launched in July 1999, is
already reaching a resolving power of $E/\Delta E \approx 1,000$ = 300
km/s in the 1 to 2 keV range,\footnotemark
\footnotetext{http://space.mit.edu/HETG/} and the Reflection Grating
Spectrometer on the XMM satellite launched in December 1999 provides
$E/\Delta E$ values from 200 to 800 in the energy range 0.35 to 2.5
keV for point sources, although the slitless design of this instrument
reduces the spectral resolution of extended sources such as clusters.
\footnotemark
\footnotetext{http://astro.estec.esa.nl/XMM/user/uhb/xmm\_uhb.html}
Typically, using multiple emission lines, one can achieve resolutions
of $10\%$ of these values, suggesting that investigations of
gravitational redshifts in rich clusters may be possible but
challenging.  Note that the thermal velocity of the ions
is smaller than $\sigma$ by a factor of $\sqrt{m_p/m_{\rm ion}}$.

\begin{figure}
\centerline{\psfig{file=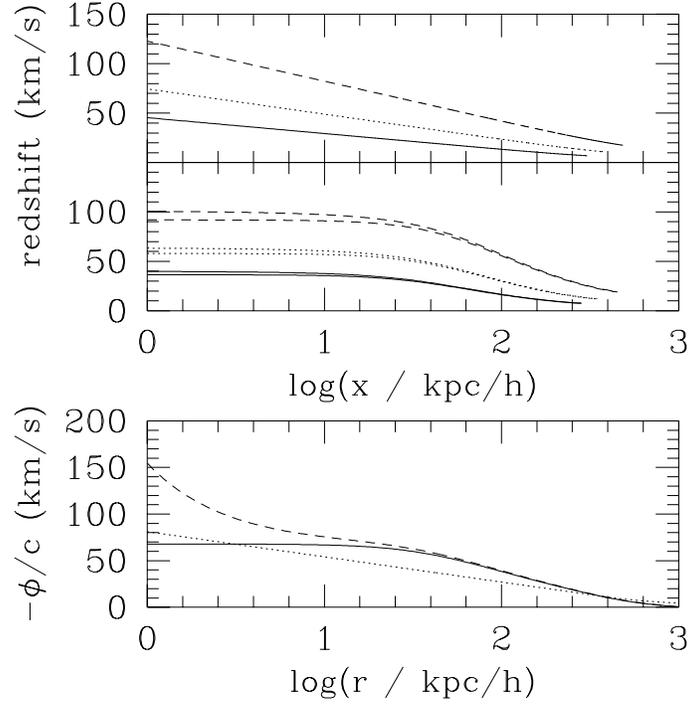,width=3.8in}}
\caption{{\em Top:}
Redshift as a function of impact parameter for a singular isothermal
profile with $M_{14} = 1.5$, $\sigma = 1000$ (solid), $M_{14} = 3$,
$\sigma = 1300$ (dotted), $M_{14} = 6$, $\sigma = 1600$ (dashed).  
{\em Center:} Redshift as a function of impact parameter for
a nonsingular isothermal profile with masses and velocity dispersions
as in the upper panel. In each pair of lines the lower line represents
the redshift purely from the profile, while the upper line includes an
additional contribution from a cD galaxy modeled as a point mass with
2\% of the cluster mass.  {\em Bottom:} Gravitational potential for
three different density profiles.  The solid lines represent the
nonsingular isothermal sphere, the dotted lines show the singular
isothermal model, and the dashed lines show a nonsingular model in
which a cD galaxy has been added.  In all cases $M_{14} = 3$, $\sigma
= 1300$, and $z_c = 0.5$ (see \S 2.1).
}
\label{fig:rofz}
\end{figure}

The next generation of satellites will significantly increase these
capabilities.  The Spectroscopy X-ray Telescope (SXT) on the proposed
Constellation X team of X-ray telescopes plans to use both
a quantum microcalorimter with 2 eV resolution
and a set of reflection gratings below 2 keV, to obtain
velocity sensitives better than 100 km/s below 1 keV and of the
order of 20 km/s at higher energies.\footnotemark
\footnotetext{http://constellation.gsfc.nasa.gov/design/resolution.html}
The narrow field detectors on the proposed XEUS observatory
in space promise similar resolutions.\footnotemark
\footnotetext{http://astro.estec.esa.nl/SA-general/Projects/XEUS/}
With such sensitivities, both core-profile studies such as shown 
in Figure  \ref{fig:rofz} and multiple-line, three-dimensional studies as 
described in \S3 will be well within reach.

Some observational issues are difficult to predict.  As line emission is
proportional to $n_{e}(r) n_{\rm ion}(r)$, metallicity gradients can bias gravitational
redshift measurements.  Such gradients have already been discovered in
the Centaurus Cluster (Fukazawa et al.\ 1994) and the cluster AWM 7 (Ezawa
et al.\ 1997), and will undoubtedly be a subject of investigation for the
current generation of X-ray satellites.  As in some cases the metallicity can 
vary by a factor of two over the cluster, these gradients must be well 
understood before our approach is applied.

Large-scale bulk motion of gas should also be detected from recent
group infall in some clusters. The disturbance of the intracluster medium by
gas infall has been claimed to be significant over a period of 3 Gyrs in
favorable conditions, before shocks are thermalised and bulk flow
through the core ceases (Roettiger, Burns, \& Loken 1996)
If infall is common then the relatively small
perturbation to the line centroid by gravity proposed here may be made
noisy by the continuous process of cluster interaction, or perhaps in
some cases relegated to a correction applied to a larger dynamical 
disturbance.  Finally the relativistic temperatures in the cluster
gas will cause a small shift in X-ray line emission by a factor of 
$\gamma-1 \approx 
\frac{k T(r)}{m_{\rm ion} c^2}$  due to relativistic beaming and
the transverse Doppler effect, but fortunately this is smaller
than the gravitational redshift by a factor of $m_p/m_{\rm ion}$.

These and other issues will await the new and next generations
of X-ray telescopes.  It is clear, however,  that 
whatever the experimental challenges, a careful measurement
of X-ray emission lines over the surface of galaxy clusters will
 be a useful probe of cluster structure and formation.

\acknowledgments

We thank Yuri Levin, Avi Loeb, and Micheal Rauch for useful conversations. TJB
acknowledges NASA grant AR07522.01-96A.


\fontsize{9}{11pt}\selectfont


\begin{deluxetable}{lrllrcl} 
\small
\tablecaption{Anomalous cD Redshifts \label{tab:anom}}
\tablewidth{0pt}
\tablehead{
\colhead{Cluster}    & \colhead{$N_{\rm gal}$} &
\colhead{Dispersion} & \colhead{$\Delta V_{cD}$} &
\colhead{$\sigma_{\rm sys}$}      & \colhead{Cooling} & \colhead{Reference} \nl
 & & \colhead{(km/s)}& \colhead{(km/s)}& \colhead{(km/s)}& \colhead{Flow} \nl
 & & & & & \colhead{($M_\odot$/yr)} \nl
}
\startdata
A193A   &32  &$760\pm 100$ & $240 \pm 150$ & 150  & --- & 
Zabludoff et al.\ (1993)
\nl         
A426    &114 &$1277 \pm 80$ & $-106 \pm  124$ & 122 & $536^{+33}_{-24}$& 
Zabludoff et al.\ (1993)
\nl         
A1644A  &82  &$940\pm 70$  & $35\pm 150$ & 113 & $11^{+40}_{-5}$ & 
Zabludoff et al.\ (1993)
\nl         
A1651   &29  &$965\pm 130$ & $210\pm 200$ & 195 & $138^{+48}_{-41}$& 
Zabludoff, Huchra, \& Geller (1990)
\nl         
A1656A  &270 &$1140\pm 50$ & $390\pm 80$ & 64 & --- & 
Zabludoff et al.\ (1993)
\nl         
A1795   &150 &$773\pm 80$  & $360\pm 150$ & 124 & $381^{+41}_{-23}$& 
Hill et al.\ (1988)
\nl         
A2199A  &68  &$820\pm 70$  & $230\pm 110$ & 104 & $154^{+18}_{-8}$& 
Zabludoff et al.\ (1993)
\nl                   
A2690   &303 &$880\pm 40$  & $440\pm 150$ & 84 & --- & 
Sharples, Ellis, \& Gray (1988) \nl  
       
\enddata
\vspace{2.0in}
\end{deluxetable}

\end{document}